\documentclass[sn-basic]{sn-jnl}

\usepackage{inputenc}
\usepackage{float}
\usepackage{graphicx}
\usepackage{algorithmicx}

\jyear{2021}%

\theoremstyle{thmstyleone}%
\theoremstyle{thmstyletwo}%

\theoremstyle{thmstylethree}%

\begin{document}
	
	\title[Observations and parameters of CR Boo]{Recent humps and superhumps observations and an estimation of outburst parameters of the AM CVn star CR Boo} 
	
	
	\author*[1]{ \sur{Daniela Boneva }}\email{danvasan@space.bas.bg}
	
	\author[2]{ \sur{Radoslav Zamanov}}
	
	\author[2]{ \sur{Svetlana Boeva}}
	
	\author[2]{ \sur{Georgi Latev}}
	
	\author[2]{ \sur{Yanko Nikolov}}
	
	\author[3]{ \sur{Zorica Cvetkovi\'{c}}}
	
	\author[4]{ \sur{Wojciech Dimitrov}}
	
	\affil*[1]{\orgdiv{Space Research and Technology Institute}, \orgname{Bulgarian Academy of Sciences}, \city{Sofia}, \postcode{1113}, \country{Bulgaria}}
	
	\affil[2]{\orgdiv{Institute of Astronomy and NAO}, \orgname{Bulgarian Academy of Sciences}, \city{Sofia}, \postcode{1784}, \country{Bulgaria}}
	\affil[3]{\orgdiv{Astronomical Observatory}, 
		\city{Belgrade}, \postcode{11060}, \country{Serbia}}
	\affil[4]{\orgdiv{Astronomical Observatory Institute, Faculty of Physics}, \orgname{Adam Mickiewicz University},
		\city{Pozna\'{n}}, \postcode{60-286}, \country{Poland}}
	
	
	
	\abstract{We present our observational results of AM CVn star CR Boo in the UBVR bands. Our observational campaign includes data obtained over 5 nights with the National Astronomical Observatory Rozhen, Belogradchik and the AS Vidojevica telescopes. During the whole time of our observations the brightness of the system varied between 13.95 - 17.23 in B band. We report the appearance of humps during the period of quiescence and superhumps during the active state of the object, (where the latter are detected in two nights). We obtain the superhumps periodicity for two nights, P{sh} = 24.76 - 24.92 min. The color during maximum brightness is estimated as  - 0.107 \textless  (B-V){0} \textless  0.257 and the corresponding temperature is in the range as 7700 [K] \textless  T(B-V){0} \textless  11700 [K]. We found that CR Boo varies from bluer to redder in the nights with outbursts activity. The star becomes bluer during the times of superhumps. }

	
	
	
	\maketitle

	\section{Introduction}\label{sec1}

	AM CVn stars are short-period binary stars in which a white dwarf accretes helium-rich material from a low-mass donor star. Their orbital periods ranges between 5 - 65 minutes \citep{2003MNRAS.340.1214P}, \citep{2010PASP..122.1133S}. 
	The AM CVn stars are rare and unusual objects. Their study can give us information about the properties of the accretion flow and the compact objects themselves. 
	By their short orbital periods and variations in photometry and spectroscopy, the AM CVn stars could also be classified as interacting binary white dwarfs or Double White Dwarf binaries (DWDs). Their main feature is that both binary components are degenerated dwarfs, the white dwarf accretes from another white dwarf companion \citep{bib31}, \citep{paczynski1967gravitational}, \citep{bib7}. 
	
	The formation of AM CVn stars currently follows two known models. The first model \citep{1979AcA....29..665T}, \citep{bib30}  is valid after common envelope phases, then the configuration consists of two degenerate dwarfs in a formation of close binary star \citep{paczynski_1976}, \citep{bib13}, \citep{1984ApJ...277..355W}. The second model describes the detached system with semi-degenerate helium star and a white dwarf companion is formed, after more than two mass-transfer phases. There is also a third, less probable channel of formation \citep{2010PASP..122.1133S}, when the donor evolves from a low-mass main sequence star \citep{2003MNRAS.340.1214P}.  
	
	The mass transfer between white dwarfs is a defining process in the evolution of AM CVn stars. Its stability or destabilization has a significant effect on the white dwarfs binary configuration \citep{bib27}. 
	When the mass transfer is in progress, while the mass ratio decreases, their orbital separation increases. At a further evolutionary point of the AM CVn objects, the direct accretion (i.e. the matter that inflow directly to the primary star or the accretor star) stops and this leads to the formation of an accretion disc. 
	The mass transfer rate changes over time, according to the binary orbital separation and also depends on the angular momentum loss in the system \citep{bib27}, \citep{bib8}. 
	
	According to some theoretical calculations \citep{1989SvA....33..606T} AM CVn stars could be a sub-part of the close binary evolution branch. Thus, they can ensure observational information to investigate the physics of helium accretion discs \citep{bib17}.
	
	It has been possible to detect the AM CVn stars by the existence mainly of helium emission lines in their spectra, since they are faint objects with an average magnitude of 12 - 17 (\cite{1987ApJ...313..757W}; \cite{1997ApJ...480..383P}; \cite{patterson1997superhumps}; \cite{bib17}).

	The AM CVn stars manifest brightness variability usually in the range of 2-4 magnitude at optical wavelengths, detected by the observational analysis (\cite{bib14}; \cite{bib16}; \cite{bib17}; \cite{bib33}) or by theoretical models \citep{1997PASJ...49...75T}. The outbursts are also reported for the AM CVn objects with orbital periods ranges from 20 - 50 min \citep{2021AJ....162..113V}.

	The AM CVn stars can be categorized by the different phases of their evolution \citep{1995Ap&SS.225..249W}, \citep{1996MNRAS.280.1035T}, \citep{bib31}: When they have short periods, their mass - transfer rates are high and the systems are in a high state. Quiescent systems have longer periods and lower mass-transfer rates.
	
	In this paper, we study the AM CVn star member CR Boo. 
	With our results of an object like CR Boo, we are contributing to improve the knowledge of the star’s color index and to enlarge the observational database. 
	We report our observational results of CR Boo in Sects. \ref{subsec3.1} and \ref{subsec3.2} and the detected observational effects are described in Sect. \ref{subsec3.3}. In Sect. \ref{sec4} the color index and the temperature are calculated at the maximum brightness. We discuss the appearance of humps and superhumps, and the variations of the parameters in Sect. \ref{sec5}. 
	
	\section{Target details}\label{sec2}
	
	CR Boo was discovered in 1986 by Palomar Green \citep{bib9} and cataloged as PG 1346+082. The observations of Wood (\cite{1987ApJ...313..757W}) show brightness variability  with amplitude 13.0 - 18.0 mag in the V band. The spectrum of CR Boo shows broad, shallow He I absorption at active state and He I emission at the quiescence state \citep{1987ApJ...313..757W}. 
	The average orbital period of CR Boo is determined as $P_{orb}=0.0170290(6)$ days \citep{1997ApJ...480..383P}, \citep{bib14}, which is about $24.5$ min $\approx 1471.3$ s. We apply these values of $P_{orb}$ in all calculations of this paper. The masses of two components were estimated to be in ranges of: $M_{1} = 0.7-1.1 M\odot$ for the mass of the primary and $M_{2} = 0.044-0.09 M\odot$ for the mass of the secondary \citep{2010PASP..122.1133S}, \citep{2007ApJ...666.1174R}.  
	
	CR Boo is an interacting double white dwarf object, in which the white dwarf primary accretes from the helium white dwarf companion \citep{paczynski1967gravitational}, \cite{bib7}, \citep{bib17}, \citep{bib32}. By its helium rich atmosphere, CR Boo is classified as DB spectral type object \citep{sion1983proposed}.  
	
	In the classification of Solheim \citep{2010PASP..122.1133S}, the AM CVn objects are divided in groups by their orbital periods and the disc\textquoteright s properties. Since the orbital period of CR Boo is in a range of $20 < Porb < 40$ min, it can be associated to the 3rd group, with a variable size of disc, producing outbursts or occasional super-outbursts. 
	
	CR Boo is categorized in the group of outburst systems \citep{bib17}, \citep{bib10} with an amplitude variations in brightness of $ > 1$ mag, lasting from days to months. 
	Two individual states of CR Boo have been observed: a faint state, with normal outbursts usually lasting one to five days \citep{2021MNRAS.502.4953D} and regular super outbursts last for several weeks \citep{bib14}, \citep{bib18};  a bright state, with frequent outburst activity, sometime prolongs for months \citep{bib18}, \citep{bib12}. The produced high outbursts during the faint state recur with a frequency in a super-cycle of about $\approx 46$ days \citep{bib16}, \citep{bib15}. Such behavior is similar to SU Uma type dwarf novae - a class of cataclysmic variables (CVs) \citep{1995Ap&SS.225..249W}.

	\section{Observations and effects}\label{sec3}
	
	\subsection{Technical details and data reduction}\label{subsec3.1}
	
	We perform $\approx 20$ hours of observations of CR Boo, distributed over 5 nights, during different observational campaigns: 1 July, 2019; 5 July, 2019; 16 April, 2020; 12 February, 2021; 4 February, 2022 \footnotemark[0]. We report observational data, obtained with four different telescopes: 
	the 2.0 m telescope of the National Astronomical Observatory (NAO) Rozhen, Bulgaria (hereafter 2m Roz), the 50/70 cm Schmidt telescope (hereafter Sch) of NAO Rozhen, the 60 cm telescope of the Belogradchik Observatory, Bulgaria and the 1.4m Astronomical Station Vidojevica (hereafter ASV) telescope, Serbia.
	
	The 2m telescope with two channel focal reductor FoReRo2 and two identical CCD cameras Andor iKON-L was used in the nights of: July 1, 2019 in V band, July 5, 2019 in UVR bands, February 12, 2021 in the B, R bands, February 4, 2022 in the UBVR bands. In the night of July 5, the 50/70 cm Schmidt telescope, equipped with CCD camera FLI PL16803 and the 1.4 m telescope at ASV, equipped with the CCD camera Andor iKON-L were used, both in B and R bands. The observations in the B band were also obtained with the 60 cm telescope of the Belogradchik Observatory (with CCD camera FLI PL16803), performed on 16 April, 2020 (hereafter 60 Bel).
	
	Data reduction was performed with standard tools for processing of CCD images and aperture photometry. Photometric standards were applied.

	Six comparison stars are chosen in the field of CR Boo (see Fig. A1 in the Appendix). Based on the B and V magnitudes published in the APASS9 catalog, the photometry of these comparison stars was performed over all available frames in these bands.
		Using r’ and i’ data from APASS9, Rc and Ic are calculated on the 3 different relations from \cite{1996AJ....111.1748F} and in the same way we obtained average magnitudes of comparison stars.
		The magnitudes in the U band for 3 standard stars are obtained using both the new (B-V) and (V-Rc) and the color indexes of normal dwarfs published by \cite{2013ApJS..208....9P}.
		The obtained stellar magnitudes of the standard stars are given in the Appendix (Table A1). 
	The stellar magnitude of CR Boo (in a given band) was obtained by ensemble aperture photometry on all field-visible comparison stars.

	\footnotetext{Note: Date formats used in this paper: YYYY-MM-DD and DD Month, Year }

	\subsection{Results from observations}\label{subsec3.2}

	During the time of our observations, the apparent magnitude of CR Boo varies in between 14.06 and 17.04 in average, in B. 
	
\begin{figure}[!htb]
		\begin{center}
			\includegraphics[width=0.75\columnwidth]{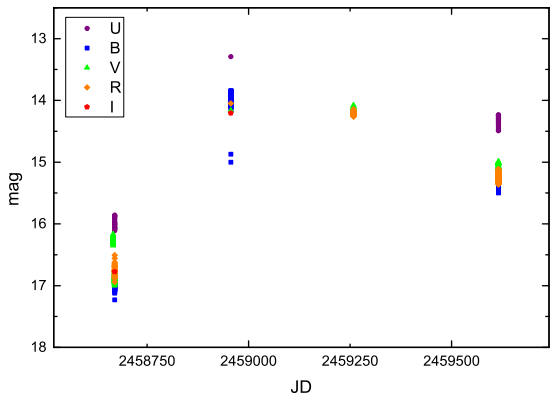}
			\caption{Light curves of CR Boo for the period of all observations, 2019 - 2022, in UBVR. The data in I band was obtained by estimations. The data were received with four different telescopes.  }
			\label{fig:1}
		\end{center}
	\end{figure}

	We see that CR Boo changes its brightness states during the time of our observations. To be more capable to assess the brightness measurements, we put all the observational data for all 5 nights in (Fig. \ref{fig:1}). 
	
	Further, we give the observational data in details, separately for each day. The journal of all observations is shown in Table \ref{tab1}.   

\begin{figure}[!htb]
		\begin{center}
			\includegraphics[width=0.70\columnwidth]{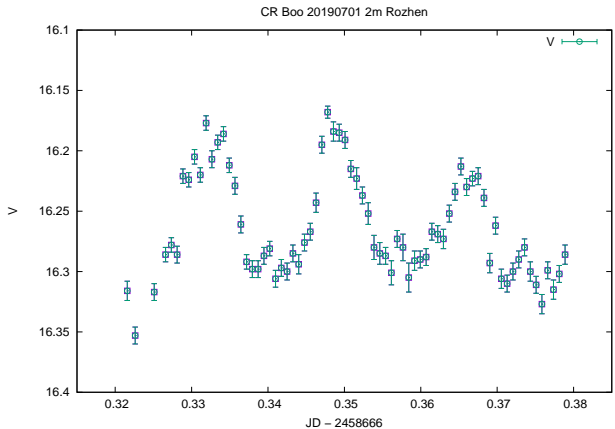} 
			\caption{Light curves of CR Boo: 1 July, 2019 in V band} 
			\label{fig:2}
		\end{center}
	\end{figure}

\begin{figure}[!htb]
		\begin{center}
			\includegraphics[width=0.70\columnwidth]{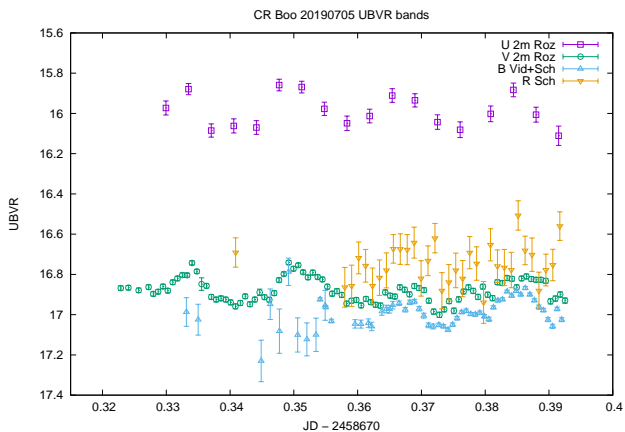}
			\caption{Light curves of CR Boo: 5 July, 2019 in UBVR bands. The data are obtained with three different telescopes }    
			\label{fig:3}
		\end{center}
	\end{figure}

	In \citep{bib4} we have reported some initial observational results from July 1, 2019 and July 5, 2019.  In the current paper, we make more precise observational analysis on these two nights. We also add data in the R band to the observations on July 5. The obtained light curves are presented in Fig. \ref{fig:2} and Fig. \ref{fig:3}. It is seen that the star\textquoteright s average brightness in V decreases with 0.7 magnitudes (Fig. \ref{fig:3}), in a period of 5 days. The average amplitude variations of the magnitude in these two nights is $\approx 0.23$ and the standard deviation varies in values of $0.04 - 0.1$.

\begin{figure}[!htb]
		\begin{center}
			\includegraphics[width=0.70\columnwidth]{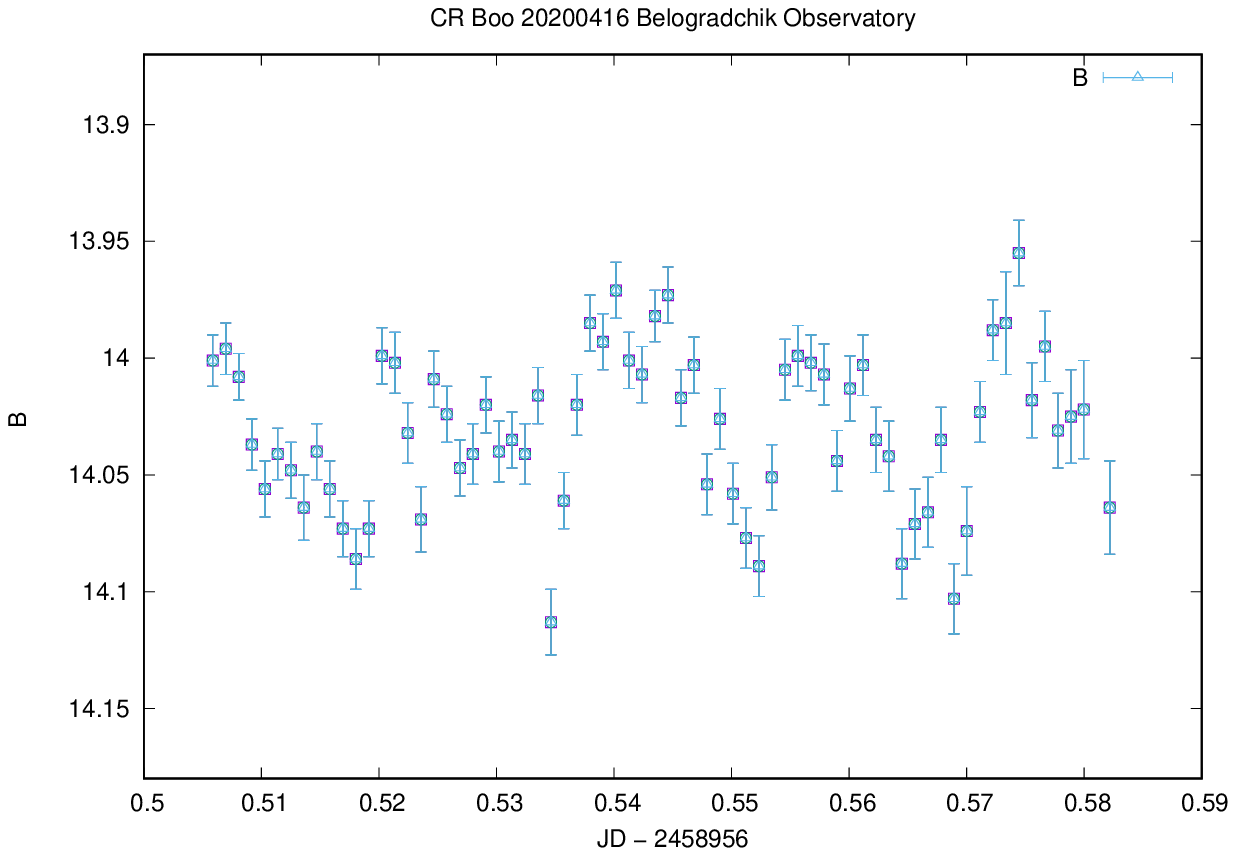}
			\caption{Light curve of CR Boo: 16 April, 2020, in B band. The data are obtained with 60 cm telescope of the Belogradchik Observatory, Bulgaria }
			\label{fig:4}
		\end{center}
	\end{figure}

	On the 3rd night (16 April, 2020), the observable brightness of CR Boo increases with 2-2.5 magnitudes in B band, comparing to the dates in July 2019. The star\textquoteright s magnitude reaches $13.95 (\pm 0.02)$ in B (Table \ref{tab1}), with $\approx 0.15$ amplitude variations and a standard deviation of $0.04$. According to the object\textquoteright s details (Section 2) and the observable magnitude of CR Boo, we can suggest that on this night the star has been in its outburst state (Fig. \ref{fig:4}). 

\begin{figure}[!htb]
		\centering
		\includegraphics[width=0.70\textwidth]{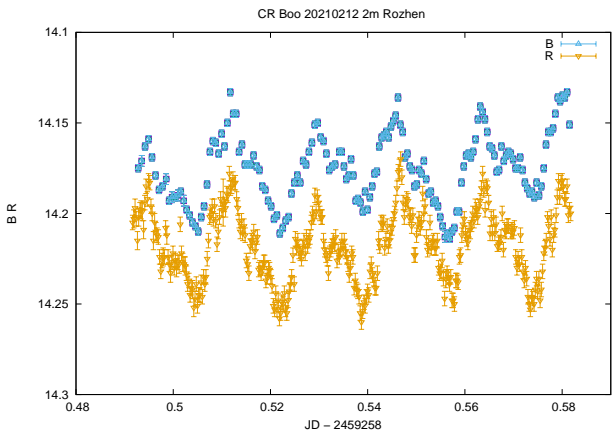}
		\caption{Light curves of CR Boo: 12 February, 2021 in BR bands. Superhumps activity are detected in both bands. The data are obtained with the 2m telescope of NAO Rozhen.  }
		\label{fig:5}
		
	\end{figure}

	The obtained observational data from 12 February, 2021 show that CR Boo is in its high state, again (Fig. \ref{fig:5}) - in comparison to the magnitudes of the nights during the July 2019 campaign. The amplitude variations of the magnitude are in a range: $0.06 - 0.08$ and the standard deviation is $0.002 - 0.02$.

\begin{figure}[!htb]
		\begin{center}
			\includegraphics[width=0.70\columnwidth]{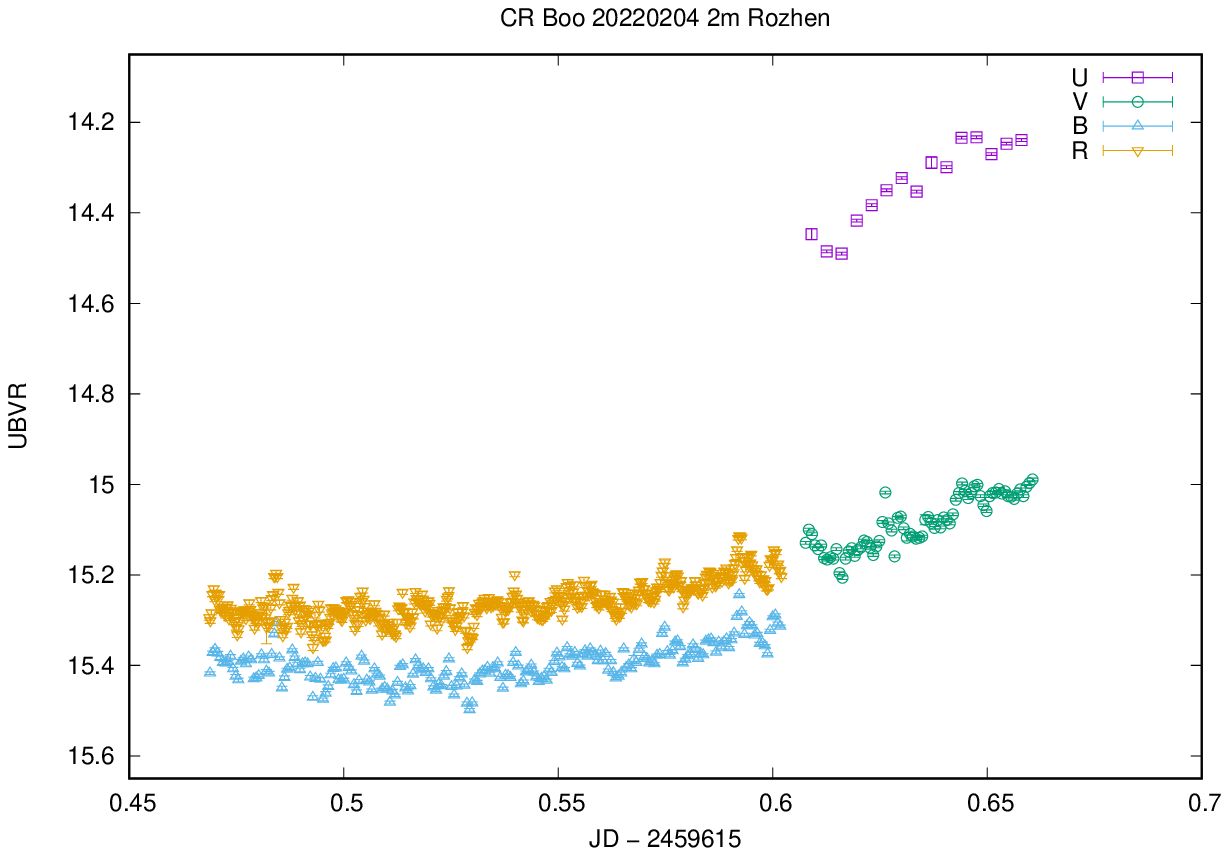}
			\caption{Light curves of CR Boo: 04 February, 2022 in UVBR bands. The data are obtained with the 2m telescope of NAO Rozhen  }
			\label{fig:6}
		\end{center}
	\end{figure}

	Our latest observations of CR Boo were performed on 04 February, 2022 (Fig. \ref{fig:6}), in UBVR bands. The magnitudes for all bands are given in Table \ref{tab1}. The amplitudes of the variations for all bands are: $0.22 - 0.26$ mag and for the corresponding standard deviations we have:  $0.009 - 0.04$. A trend of increase in brightness is seen, more clearly expressed in U and V bands in the end of the night. 
	
	\begin{sidewaystable}[!h]
		\sidewaystablefn%
		\begin{center}
			\caption{Journal of observations of CR Boo in UBVR} \label{tab1}
			\begin{tabular}{@{}cccccccc@{}}
				\toprule
				
				Date  &  Band  & Telescope  & Exp.time [s] & start / end [UT] & max/min [mag] & Avr [mag] & Error  \\
				
				\hline
				
				2019-07-01  & V & 2m Roz & 60 & 19:41 / 21:03 & 16.17 / 16.35 & 16.26 & $\pm 0.01$  \\ 
				
				\hline
				2019-07-05  &  U &  2m Roz & 300 & 19:50 / 21:19 & 15.85 / 16.11 & 15.98 &  $\pm 0.04$\\  
				&  B &  50/70 Sch & 120 & 20:15 / 20:30 & 16.78 / 17.23 & 17.04 & $\pm 0.1$  \\
				&  B &  Vid  & 30/60 & 20:28 / 21:22 & 16.87 / 17.08 & 16.98 & $\pm 0.02$  \\
				
				&  V & 2m Roz & 60  & 19:42 / 21:22 & 16.74 / 17.00 & 16.89 & $\pm 0.02$ \\
				&  R &  Sch & 90 & 20:08 / 21:22 & 16.51 / 16.93 & 16.74 & $\pm 0.1$  \\
				\hline
				2020-04-16  & B & 60 Bel & 90 & 23:58 / 02:05 & 13.95 / 14.15 & 14.06 & $\pm 0.02$ \\
				
				\hline
				
				2021-02-12 &  B &  2m Roz & 60/50/40 & 23:55 / 02:03 & 14.13 / 14.21 & 14.17 & $\pm 0.01$ \\ 
				&  V &  2m Roz  & 40 & 02:26 / 02:40 & 14.08 / 14.14 & 14.12 & $ \pm 0.005$ \\
				&  R &  2m Roz  & 20 & 23:54 / 02:04 &  14.14 / 14.26 & 14.22 & $\pm 0.01$ \\  
				
				\hline
				
				2022-02-04 & U & 2m Roz & 300 & 02:40 / 03:56 & 14.23 / 14.49 & 14.33  & $\pm 0.01$ \\
				&       B & 2m Roz & 50 & 23:20 / 02:32 & 15.24 / 15.49 & 15.39 & $\pm 0.01$ \\
				&       V & 2m Roz & 60 & 02:40 / 03:56 & 14.98 / 15.21 & 15.08 & $\pm 0.005$ \\
				&       R & 2m Roz & 20 & 23:20 / 02:32 & 15.11 / 15.36 & 15.26 & $\pm 0.01$ \\          
				\botrule
			\end{tabular}
		\end{center}
	\end{sidewaystable}

	\subsection{Observational effects. Humps and superhumps}\label{subsec3.3}
	
	During the times of our observations, two observational effects are clearly distinguished.  They appeared as short-period, low-magnitude brightness variations. The authors recognized them as humps and superhumps \citep{bib14}, \citep{bib17}. The humps are observed in the quiescence low state of the cataclysmic variables and AM CVn stars. They appear with a periodicity $P_{h}$ similar to the binary orbital period. On the other hand, the superhumps\textquoteright periodicity $P_{sh}$ is a few percent longer than the binary period and they can be observed during the outbursts state of the objects \citep{1995Ap&SS.225..249W}.
	
	Using the observational data of CR Boo (Section 3.2), we obtain the periodicity of the maxima in brightness variations for each observational night. To analyze these periodicity, we apply the PDM (Phase Dispersion Minimization) method by \cite{1978ApJ...224..953S}. We also check our results with additional software packages like an OnLine based PGRAM 
	(exoplanetarchive.ipac.caltech.edu) and PerSea \citep{2005BaltA..14..205M}) based on fast and statistically optima period search in an uneven sampled observation method by A. Schwarzenberg-Czerny  \citep{1996ApJ...460L.107S}.

	The measured period of the amplitude variations on July 1 and July 5 (see Table \ref{tab3}) is approximately the same (with a difference of $\pm 0.07 - 0.2$ min) as the orbital period of $24.5$ min.

	In the previous section, we defined that on the first two nights the object was in its quiescent state. Following the above terminology, these periodic small-scale amplitude variations in the magnitude of CR Boo could be related to the manifestation of humps, more clearly seen in the U and V bands. The humps are also called orbital humps, since they usually appear regularly with the orbital period.   
	
	On the second two nights, 16 April, 2020 and 12 February, 2021, CR Boo is in its outburst state (see Subsection 3.2). The estimated average periodicity of the maximum brightness (see Table \ref{tab3}) on 2020-04-16 and on 2021-02-12 are slightly ($\approx 1.5 \% $) higher than the orbital period of the binary. 
	The observed brightness variations then are assumed to be an appearance of superhumps. These values are close to the estimated superhumps periods of CR Boo in the analysis of \cite{patterson1997superhumps} and \cite{bib14}. 
	
	On the last date, 04 February 2022, the star was in a condition with a rising brightness during the night. In a frame of $\approx 90$ minutes its magnitude increased with  $\approx 0.26$ mag in the U and V bands. It was probably transitioning to an outburst state. Brightness variations with small-amplitudes of $0.023 - 0.110$ mag are observed in the B and R bands. These brightness variations have very short periodicity, in a range of $4.42 - 10.30$ min. We found they look much more like quasi-periodic oscillations in a stage before the star is turning to the higher state.

	\begin{table*}[!h]
		\begin{center}
			\caption{Periodicity of the maximum brightness variations, calculated for each observational nights. } \label{tab3}
			\begin{tabular}{@{} ccc @{}}
				\toprule
				
				Parameter    & $P_{h}$ [min] & $P_{sh}$ [min]    \\
				Date     &              &                       \\
				\hline \\                    	    
				
				2019-07-01  &  $23.41-24.40$  & -     \\
				&  $(\pm 0.05)$   &         \\
				\hline  \\
				2019-07-05 & $ 23.71-24.60 $  & -    \\
				&    $(\pm 0.035 )$  &        \\
				\hline \\
				2020-04-16  &      -          & $24.76$  \\
				&          &  $(\pm 0.023)$ \\
				\hline \\
				2021-02-12  &    -          & $24.92$    \\
				&          &  $(\pm 0.0012)$  \\
				
				\hline \\       
				2022-02-04  & $4.42-10.30$  &  -  \\
				&        $(\pm 0.45)$   & -  \\
				
				\hline  \\
				Ref. values [min]  & $23.21$ \footnotemark[1] & $(24.79-25.44)$ \footnotemark[2]  \\
				&                          &                                          $24.78$ \footnotemark[3] \\
				
				\botrule
			\end{tabular}\\
			\footnotetext{Note: Ref. Table 3}
			\footnotemark[1]{\citep{bib4}}
			\footnotetext{Note: Ref. Table 3}
			\footnotemark[2]{\citep{bib14}}
			\footnotetext{Note: Ref. Table 3}
			\footnotemark[3]{\citep{patterson1997superhumps}}
			
		\end{center}
	\end{table*}

	\section{System parameters: Color index and temperature }\label{sec4}

	The observations in UBVR filters on July 5, April 16th, February 12th and February 4 allow us to estimate the color indexes of CR Boo. Following the data, we obtain the indices values at average brightness of these four nights.

	The derredened color indexes $(B-V)_{0}$  are obtained using  the color excess  $E(B-V) = 0.013 \pm 0.006$, calculated on the base of the field - averaged selective extinction $<Av> = 0.04 \pm 0.02$ \citep{2007ApJ...666.1174R}, excluding the negative results for the Ref.3 star for CR Boo and the standard extinction law  \citep{1989ApJ...345..245C}, \citep{1999PASP..111...63F}, \citep{2005ApJ...619..931I}. 
	
	The observed color index on July 5 is slightly larger than zero or $\approx 0$, which shows a tendency for the source to become red at maximum brightness. The usually rare observations for CR Boo in U band, give negative value of the color index $(U-B)_{0}$ for both nights: 2019-07-05 and 2022-02-04. 
	
	Further, using the derredened color index $(B-V)_{0}$, we could calculate the color temperature.  The formula of \citep{bib1} is appropriate to apply:

	\begin{equation}
		T[K] = 4600 [{\frac {1}{0.92(B-V)_{0}+1.7}}+{\frac {1}{0.92(B-V)_{0}+0.62}}]      
	\end{equation}

	The obtained value for the temperature on July 5th is then: 
	$T(B-V)_{0} \approx 8900 \pm 400 K$
	
	The B-V index on 16 April, 2020 goes from zero to a slightly negative value, which is an indication of a bluer and hotter source of the superhumps events on this night. This results on the temperature variations, which reaches the value at maximum brightness: $T(B-V)_{0}\approx 11700 \pm 400 K$. 
	
	The observations during the February 12th, 2021 campaign show that the source of the appeared superhumps is rather redder in relation to the B and U color. Based on the B-V index on the 2021-02-12, the estimation of the average temperature gives a lower value: 
	$T(B-V)_{0} \approx 9600 \pm 200 K$, which is in accordance with the reddening of the source.

	The lightcurves obtained on February 4, 2022 displays the observable trend in brightness. These excursions are typical for CR Boo and were described by \cite{patterson1997superhumps}. For that reason, in order to estimate the color indexes cited above, we use the average trend values in UBVR bands. The obtained temperature is then: 
	
	$T(B-V)_{0} \approx 7700 \pm 220 K$.

	The obtained average values of the color indices for each date and the corresponding color temperatures are given in Table \ref{tab4}.

	\begin{table*}[!h]
		\begin{center}
			\caption{Color index and color temperature for the four observational dates.} \label{tab4}
			\begin{tabular}{@{}ccccc@{}}
				\hline
				
				Date /  &    2019-07-05 & 2020-04-16  & 2021-02-12 & 2022-02-04 \\
				Parameter &   \\
				\hline

				&                   &                     &  & \\
				$U-B$ & $- 1.09 \pm 0.04$ & $- 0.772 \pm 0.04$ & - & $- 1.06 \pm 0.04$\\
				\hline
				$(U-B)_{0}$ &  $- 1.097 $ & $ - 0.782 $ & - & $- 1.071 $\\
				&  $(\pm 0.04)$ &  $(\pm 0.04)$  & - & $(\pm 0.05)$\\
				\hline
				$B-V$ & $0.12 $ & $- 0.094 $ & $0.054 $ & $0.27 $\\
				&  $(\pm 0.02)$ &  $(\pm 0.023)$  & $(\pm 0.021)$ & $(\pm 0.03)$\\
				\hline
				$(B-V)_{0}$ &   $0.109$  & $ - 0.107 $ &  $0.041$  & $0.257$ \\
				&  $(\pm 0.04)$ &  $(\pm 0.03)$  & $(\pm 0.02)$ & $(\pm 0.04)$\\
				
				\hline  
				$B-R$ & $0.27 \pm 0.03$ & - & $-0.042 \pm 0.021$ & $0.13 \pm 0.01$ \\
				
				\hline
				$ V-R$ & $0.15 \pm 0.05$ & $0.110 \pm 0.05$ & - & $-0.15 \pm 0.02$ \\
				\hline
				$T_{col}(B-V)_{0} [K]$ & $8900$    &  $11700$ & $9600$ & $7700$ \\
				&           $(\pm 400)$  &     $(\pm 400)$   &  $(\pm 200)$ & $(\pm 220)$ \\
				
				\hline 
				
				\\
				
				\hline
			\end{tabular}
		\end{center}
	\end{table*}

\begin{figure}[!htb]
		\begin{center}
			\includegraphics[width=0.60\textwidth]{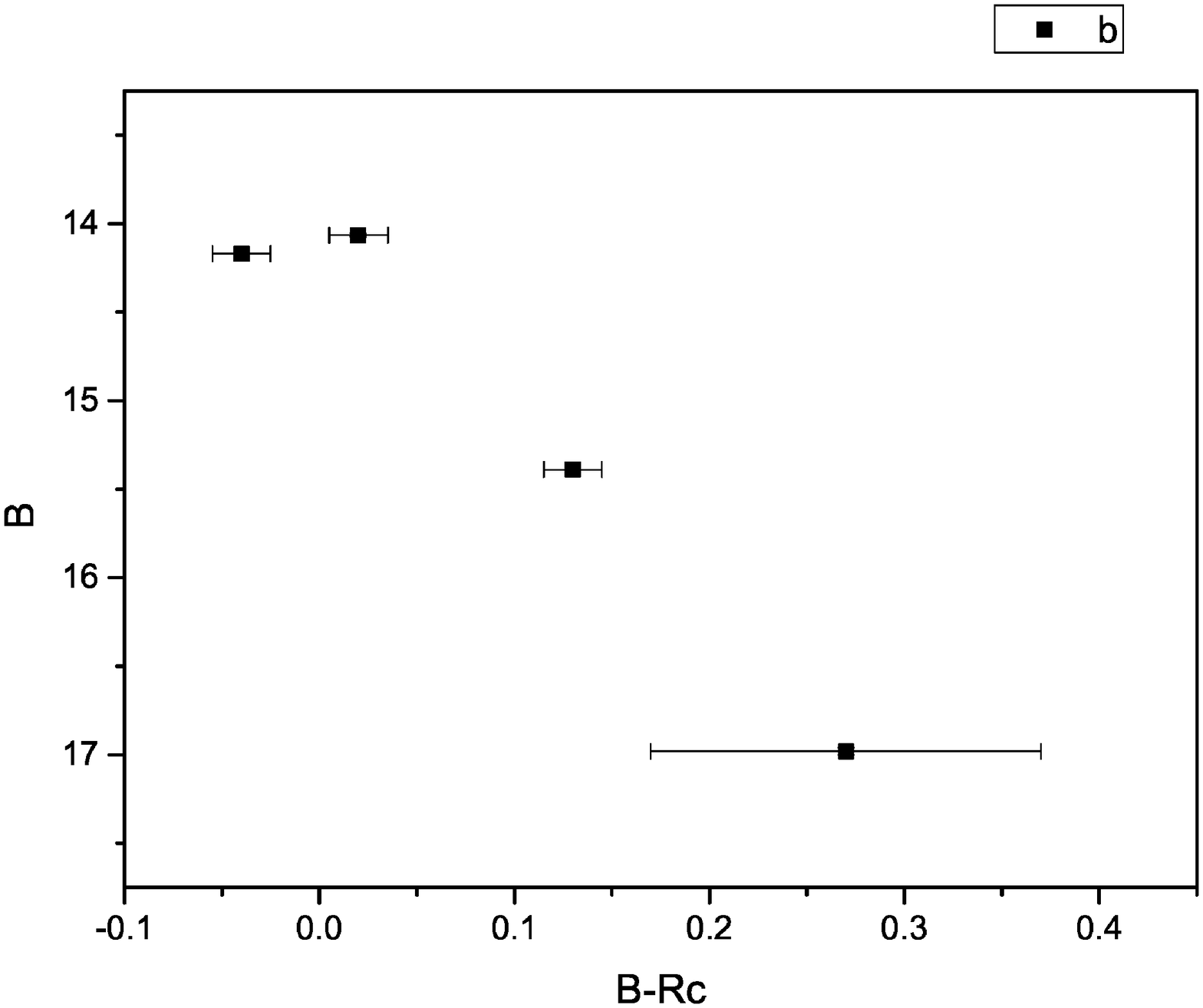} 
			\hfill
			\includegraphics[width=0.55\textwidth]{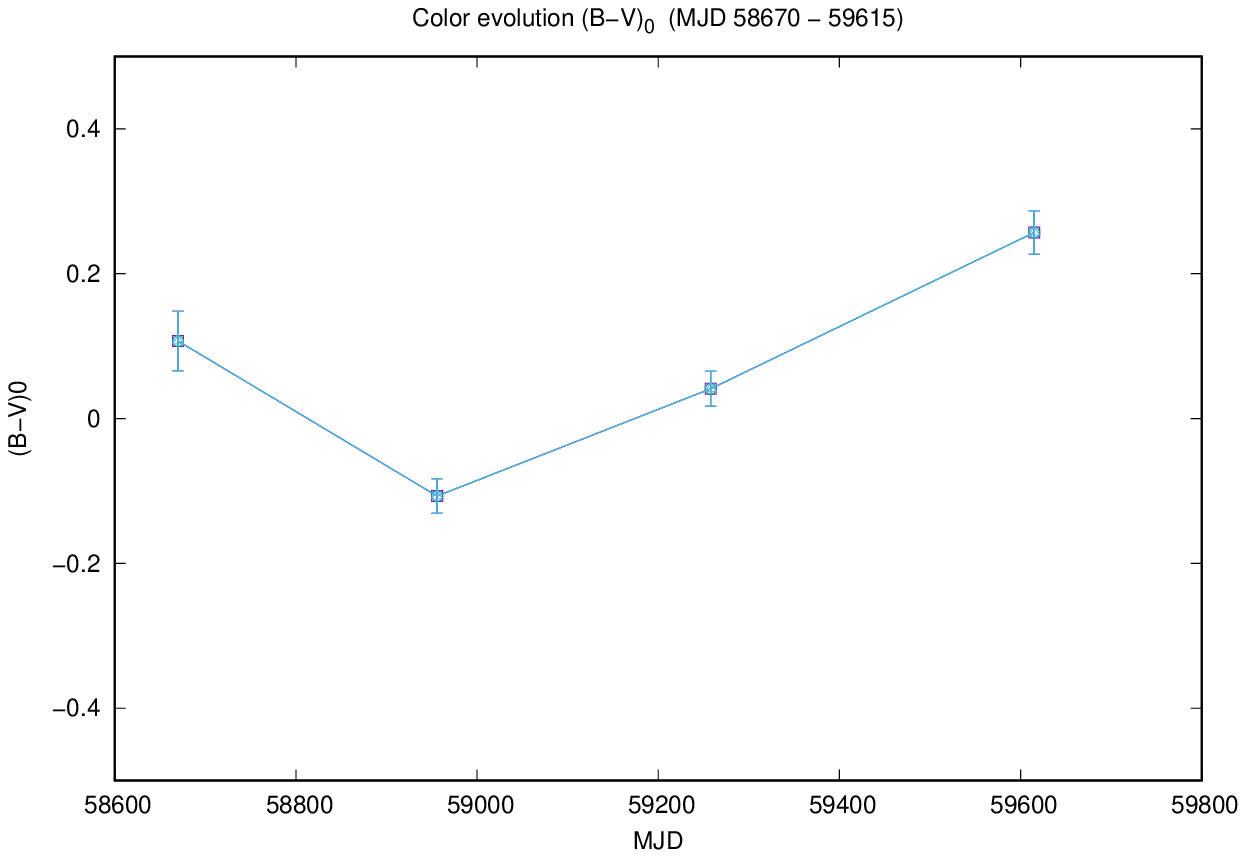}
			\hfill
			\includegraphics[width=0.55\textwidth]{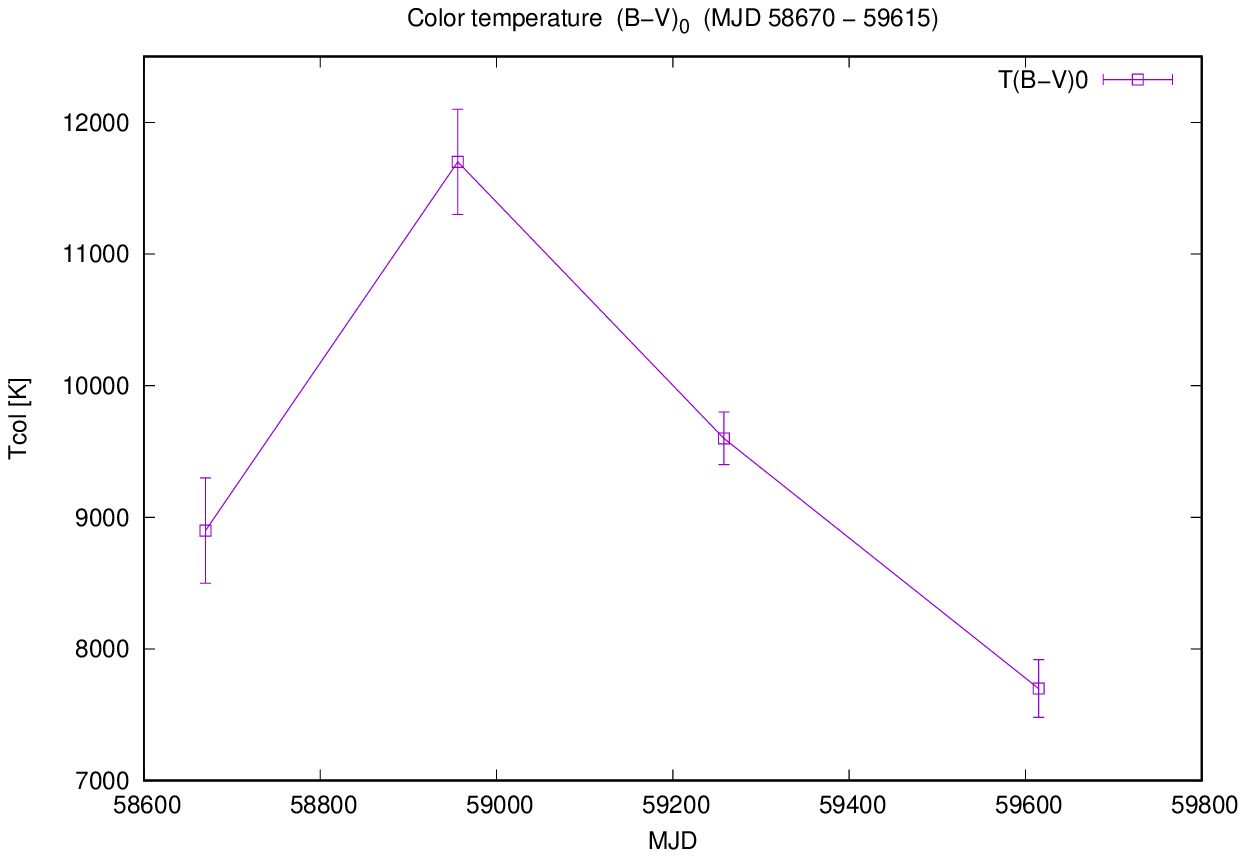}
			
			\caption{The color – magnitude diagram B: B-R (upper) presents CR Boo in outburst state. It is seen, the star becomes bluer and brighter. The color evolution $(B-V)_{0}$ (middle) and temperature variations (lower) for the period from 2019-07-05 to 2022-02-04, in MJD.     }
			\label{fig:7}
		\end{center}
	\end{figure}

	In Fig. \ref{fig:7} we present the color evolution and the temperature gradient through the different states of CR Boo for the time of observations. The figure shows their variations during the humps and superhumps activity (see also Fig. \ref{fig:1} ).  
	
		\section{Discussion}\label{sec5}
		
		AM CVn stars are objects in the final stage of binary stars evolution. They create a medium to study the physical properties of such systems, to obtain more information about the white dwarf stars and to understand the double white dwarfs evolution, respectively.

		\subsection{Humps and superhumps production}\label{sec5.1}

		As we have seen in Section 3, the observational results show manifestation of small scale amplitude semi-periodic variations during the low state of the CR Boo on the 1st two days of observations (2019-07-01 and 2019-07-05). The similar variations, with a longer period, are observed in the third and fourth nights, when the star is in its outburst state (2020-04-16 and 2021-02-12). Here, we discus the probable sources of their production and their probable positions throughout the accretion disc.
		
		Following the results in Sects.  \ref{subsec3.2} and \ref{subsec3.3}, our first assumption is that the humps during dates 2019-07-01 and 2019-07-05 are an exhibition of the orbital humps \citep{2002A&A...383..574O}.   
		The acceptance is that the source of those humps is the periodic appearance of the hot spot, placed in the outer part of the accretion disc. While the CR Boo binary system rotates, the periodical disturbances in the light curve are produced, when the hot spot is at the position facing the direction of observations. This happens in every full rotational cycle and CR Boo is in its quiescence state.

		From the geometrical point of view, the orbital inclination of CR Boo, $i = 30^{\circ}$, \citep{2001A&A...373..222N} could effect on the observable to us line of sight. 
			In the result, by all this configuration a partially effect on the light curve could be detected. This might be the minor disturbances in the brightness, or humps.

		 It is known that the superhumps during the outburst\textquoteright s activity could be produced by a tidal instability, with an effect of the disc precession (see \citep{bib11},  \citep{bib19}, \cite{2011ApJ...741..105W}, \citep{bib21}, \citep{1995Ap&SS.225..249W}).

		The production of superhumps could also be caused by other mechanisms. As a result of interaction, the spiral density wave formation on the outer disc edge could have an significant effect on the light curve and correspondingly on the superhump\textquoteright  production (\citep{1998ApJ...506..360S}, \citep{bib22}). Then, the brightness increases by the energy releases by the density strengthening at the places of spiral-density wave interactions with other disc\textquoteright s shock formations.     
		
		Along with the mention above, if a blob exists onto the accretion disc surface, its role can be pointed here as to its interaction with spiral arms, which leads to the production of short-time oscillations on the light curve.   
		
		On the other hand, the tidal wave coming from the secondary star through the Lagrange point $L_{1}$ could make the hot spot parameters (as its size and density) unstable. This may cause further unregular or fading humps production.

	\subsection{Variations in parameters}\label{sec5.2}
		
		According to the calculated indices (section \ref{sec4}), the color of the star varies from U, B to R, in different states and dates. It becomes ultraviolet and the same time red (on 2019-07-05), when it is in a low state, with humps activity. When CR Boo is in a low to rising state (on 20220204), with a manifestation of quasi-periodic oscillations, an ultraviolet excess is observed and the star stays redder. We have two different colors states of the star, both during the outbursts on two different nights (2020-04-16 and 2021-02-12).    
		
		Comparing the results between two of the observational dates (2020-04-16 and 2021-02-12), when the superhumps are observed, it is notable that: on the first night (2020-04-16) the object is bluer and its color temperature on this night is a bit higher by $\approx 2100 \pm 450 K$ than the value of the second night (Table \ref{tab4}).
		
		A source of these variations in the temperature in B band has been probably changed its parameters at the later date (2021-02-12), when the star is redder and we have the lower values of the temperature.

		Many Cataclysmic variables staying in a brightness low states have analogous behavior of their color indices – e.g. MV Lyr \citep{1981ApJ...251..611R}, KR Aur \citep{2006ASPC..349..197B} etc. When they became in a very low state (such as a deep state or a deep minimum), they show a higher temperature and a bluer color, compared to the high state. This could be caused by the appearance in the energy distribution of the white dwarf or the inner hot parts of a weak accretion disk. 
		Comparing CR Boo, by our results, with other AM CVn objects \citep{2021MNRAS.505..215R}, we see the star is in contrast to the behavior of binary SDSS 0807 and it is similar to AM CVn SDSS 1411, on the nights of 2019-07-05 and 2020-04-16. On the night of 2021-02-12, CR Boo\textquoteright s color is in a close consistent with SDSS 1137 and SDSS 0807. The behavior of our object on 2022-02-04 looks more like SDSS 1411.  
		
		In the case of our observations, CR Boo generally gets redder as its brightness weakens, except in the night of February, 12, 2022. This is probably due to the disc is being sufficiently bright, even in the lowest states, that we have observed and it is dominating in the system\textquoteright s emission.
		
		The resulting negative value of the color index $U-B < 0$ during our observations is an indication of a hotter radiation zone in the accretion disc around the primary star. The heating parts of accretion disc could also be responsible for the humps and superhumps productions.

		\section{Conclusion}\label{sec6}
		
		We presented our observational results of AM CVn star CR Boo in the UBVR bands, performed over five nights during different observational campaigns. The data were obtained with the Rozhen National Astronomical Observatory, Belogradchik and AS Vidojevica telescopes. The brightness of the system varied between $13.95 - 17.23$ in B band. We confirmed the appearance of humps during the quiescence state and superhumps during the active state in the CR Boo type of objects. The superhumps periodicity is obtained, $P_{sh} \approx 24.76 - 24.92$ min, for the nights, when the object was in its probable outburst state. During the superhumps, the color was $ -0.094 < B-V < 0.054$. We calculated the color temperature using the dereddening color indices and it is in the range $9600 [K] < T(B-V)_{0} < 11700 [K]$. We observe an increasing temperature during the superhumps observation period, compared to the values of the humps period with an average of $2100 \pm 450 K$ for both nights. The star becomes bluer when it is brighter in times of superhumps. 
		We found that during the two nights with superhumps activity, CR Boo has different behavior: the star became blue in the first night, but it is red in the second night.

		\bmhead{Acknowledgments}
		The authors thank the anonymous referee for the useful comments that helped us to improve the paper. 
		The authors thank to the grant: Binary stars with compact objects, $K\Pi -06-H28/2$   $08.12.2018$ (Bulgarian National Science Fund). 
		This research has made use of the NASA Exoplanet Archive, which is operated by the California Institute of Technology, under contract with the National Aeronautics and Space Administration under the Exoplanet Exploration Program.
		
		\section*{Declarations}

		\begin{itemize}
			\item This work was supported by the grant [Binary stars with compact objects, $K\Pi -06-H28/2$   $08.12.2018$ (Bulgarian National Science Fund)].
			\item The authors declare that they have no conflicts of interest. 
			\item Ethics approval 
			\item All authors contributed to the study conception and design. Observations were performed by [Svetlana Boeva], [Daniela Boneva], [Georgi Latev], [Yanko Nikolov] and [Zorica Cvetkovi\'{c}]. Material preparation, data collection and analysis were performed by [Svetlana Boeva], [Daniela Boneva], [Radoslav Zamanov] and [Georgi Latev]. The text preparation and correction by [Daniela Boneva],[Svetlana Boeva],[Georgi Latev]and [Wojciech Dimitrov]. The first draft of the manuscript was written by [Daniela Boneva] and all authors commented on previous versions of the manuscript. All authors read and approved the final manuscript.
		\end{itemize}
		
		\begin{appendices}
			
			\section{}\label{secA}
			
			In this Appendix, we present information about the comparison stars, used in the photometric analysis of the recent observations of CR Boo (Section 3.1). The chart in Fig. A1 shows the configuration of the stars in the field of CR Boo, where the comparison stars are denoted with numbers and blue circles. The obtained magnitudes of the standard stars are shown in Table A1.
			
			\begin{figure}[!htb]
				\begin{center}
					\includegraphics[width=0.75\columnwidth]{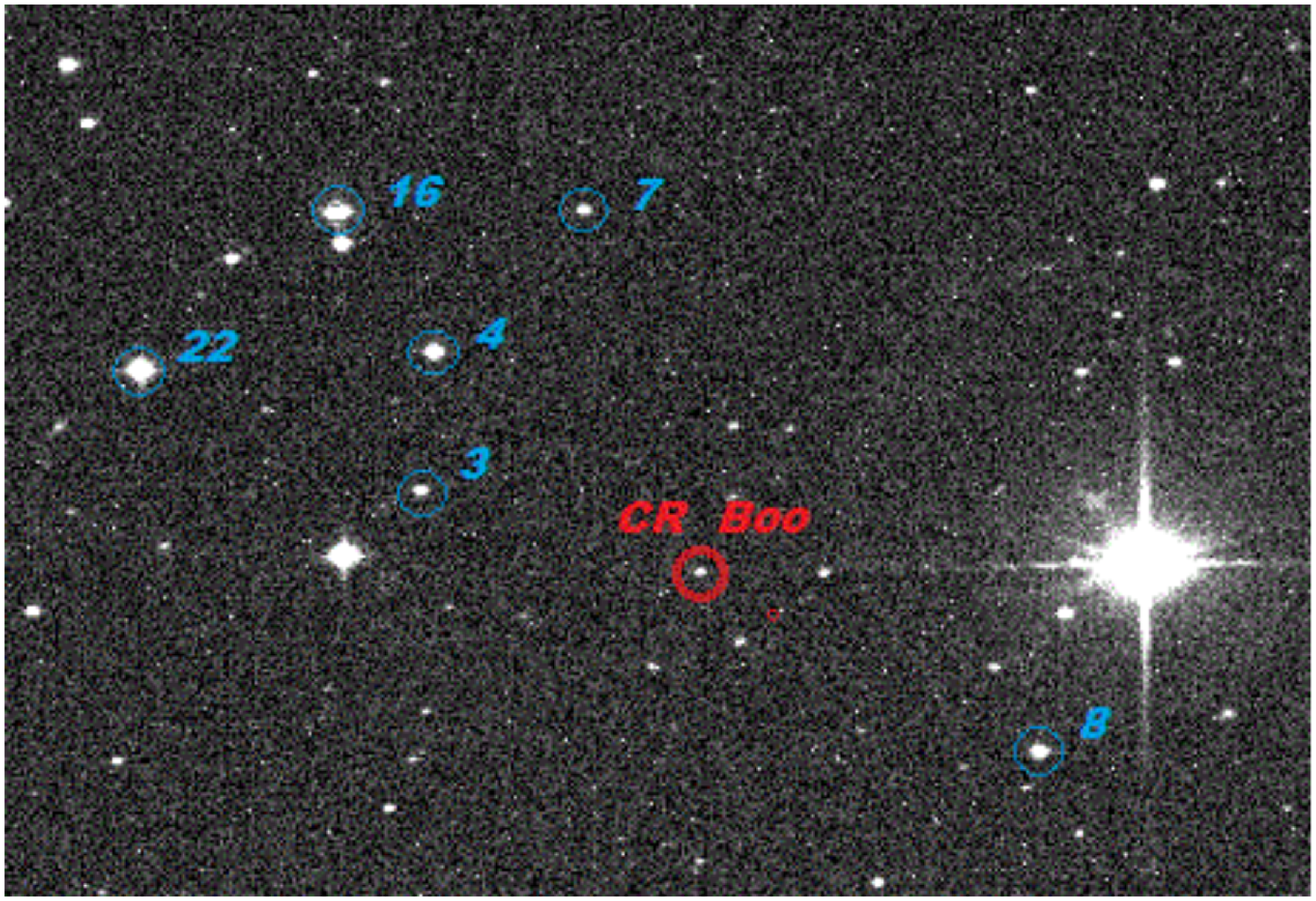}
					\caption{A chart of comparison stars. The stars are denoted with blue circles and numbers} 
					\label{fig:8}
				\end{center}
			\end{figure}

			\begin{table}[!htb]
				\caption{Standard stars photometry (Johnson - Cousins). The stars numbered refer to those in Fig. 8} \label{tab5}
				\begin{tabular}{@{}c c c c c c @{}}
					
					\toprule
					Star / Band & U & B & V & R & I \\
					\toprule
					\\
					(3) & - & $16.377$ $\pm 0.020$   &  $15.545$  $\pm 0.010$ & $15.088$ $\pm 0.010$ & $14.628 \pm 0.05$   \\ 
					\hline                     	    
					\\
					(4) & $16.30$ $\pm 0.05$ & $14.959$ $\pm 0.030$ & $13.742$ $\pm 0.025$ & $12.991$ $\pm 0.020$ & $12.373 \pm 0.005$  \\
					\hline
					\\
					(7) & - & $16.335$ $\pm 0.030$ & $15.758$ $\pm 0.010$ & $15.399$ $\pm 0.010$ & $15.038 \pm 0.05$  \\
					\hline
					\\
					(8) & - & $14.911$ $\pm 0.020$ & $14.152$ $\pm 0.015$ & $13.762$ $\pm 0.030$ & $13.357 \pm 0.02$ \\
					\hline
					\\
					(16) & $14.07$ $\pm 0.03$ & $13.229$ $\pm 0.015$ & $12.259$ $\pm 0.015$ & $11.670$ $\pm 0.005$ & $11.147 \pm 0.01$ \\
					\hline
					\\
					(22) & $12.64$ $\pm 0.02$  & $12.621$ $\pm 0.015$ & $12.070$ $\pm 0.005$ & $11.781$ $\pm 0.010$ & $11.472 \pm 0.005$ \\
					
					\\  
					\botrule  
					
				\end{tabular}
			\end{table}
			
		\end{appendices}

		\bibliography{Bibliographyour}
		
		
	\end{document}